\documentclass[fleqn,twoside]{article}
\usepackage{espcrc2}

\usepackage{graphicx}
\usepackage[figuresright]{rotating}

\newcommand{\AmS}{{\protect\the\textfont2
  A\kern-.1667em\lower.5ex\hbox{M}\kern-.125emS}}

\def\hMpc{\,h^{-1}{\rm Mpc}}
\def\alt{\raise0.3ex\hbox{$\;<$\kern-0.75em\raise-1.1ex\hbox{$\sim\;$}}}
\def\agt{\raise0.3ex\hbox{$\;>$\kern-0.75em\raise-1.1ex\hbox{$\sim\;$}}}

\newcommand{\gtrsim}{\; ^> \!\!\!\! _\sim \;}
\newcommand{\url}{\texttt}
\newcommand{\be}{\begin{equation}}
\newcommand{\ee}{\end{equation}}
\newcommand{\bea}{\begin{eqnarray}}
\newcommand{\eea}{\end{eqnarray}}

\title{The signature of local cosmic structures on the ultra-high energy cosmic
ray anisotropies}

\author{A. Cuoco\address[MCSD]{Dipartimento di Scienze Fisiche, Universit\`{a} di
Napoli ``Federico II", and INFN-Sezione di Napoli, \\ Complesso
Universitario di Monte Sant'Angelo, Via Cintia, I-80126 Napoli,
Italy}%
\thanks{electronic address: cuoco@na.infn.it}}

\begin{document}

\begin{abstract}
Current experiments collecting high statistics in ultra-high
energy cosmic rays (UHECRs) are opening a new window on the
universe facing the possibility to perform UHECR astronomy. Here
we discuss a large scale structure (LSS) model for the UHECR
origin for which we evaluate the expected large scale anisotropy
in the UHECR arrival distribution. Employing the IRAS PSCz
catalogue as tracer of the LSS, we derive the minimum statistics
needed to reject or assess the correlation of the UHECRs with the
baryonic distribution in the universe, in particular providing a
forecast for the Auger experiment. \vspace{0pc}
\end{abstract}

\maketitle
%%%%%%%%%%%%%%%%%%%%%%%%%%%%%%%%%%%%%%%%%%%%%%%%%%%%%%%%%%%%%%%%%%%%%%%
\section{Introduction}
%%%%%%%%%%%%%%%%%%%%%%%%%%%%%%%%%%%%%%%%%%%%%%%%%%%%%%%%%%%%%%%%%%%%%%%
At energies above a few~$\times 10^{19}\,$eV, which we will refer
to as the ultra-high energy (UHE) regime, protons propagating in
the Galaxy retain most of their initial direction. Provided that
the extragalactic magnetic field (EGMF) is negligible, UHE protons
will therefore allow us to probe into the nature and properties of
their cosmic sources. However, due to quite steep CR power
spectrum, UHECRs are extremely rare (a few particles km$^{-2}$
century$^{-1}$) and their detection calls for the prolonged use of
instruments with huge collecting areas. One further constraint
arises from an effect first pointed out by Greisen, Zatsepin and
Kuzmin~\cite{Greisen:1966jv,Zatsepin:1966jv} and since then known
as GZK effect: at energies $E\gtrsim 5\times 10^{19}$ eV the
opacity of the interstellar space to protons drastically increases
due to the photo-meson interaction process $p+\gamma_{\rm
CMB}\to\pi^{0(+)}+p(n)$ which takes place on cosmic microwave
background (CMB) photons. In other words, unless the sources are
located within a sphere with radius of ${\cal O}$(100) Mpc, the
proton flux at $E\gtrsim 5\times 10^{19}$ eV should be greatly
suppressed. However, due to the very limited statistics available
in the UHE regime, the experimental detection of the GZK effect
has not yet been firmly established.

A detailed knowledge of the UHECR framework is still missing. Only
recently, for instance, magnetic fields were included in
simulations of large scale structures
(LSS)~\cite{Dolag:2003ra,ems}. Qualitatively the simulations agree
in finding that EGMFs are mainly localized in galaxy clusters and
filaments, while voids should contain only primordial fields.
However, the conclusions of Refs.~\cite{Dolag:2003ra}
and~\cite{ems} are quantitatively rather different and it is at
present unclear whether deflections in extragalactic magnetic
fields will prevent astronomy even with UHE protons or not.
Another large source of uncertainty is our ignorance on the
chemical composition of UHECRs, mainly due to the need to
extrapolate for decades in energy the models of hadronic
interactions, though important progress are expected from new high
quality data and deconvolution techniques  (see e.g.
\cite{Antoni:2005wq}). Future accelerator measurements of hadronic
cross sections in higher energy ranges will also ameliorate the
situation, but this will take several years at least.

From now on, therefore, we shall work under the assumptions that
UHE astronomy is possible, namely: i) proton primaries; ii) EGMF
negligibly small; iii) extragalactic astrophysical sources are
responsible for UHECRs acceleration. A possibility favoring these
hypotheses is that relatively few, powerful nearby sources are
responsible for the UHECRs, and the small scale clustering
observed by AGASA~\cite{Takeda99} may be a hint in this direction.
However, the above quoted clustering has not yet been confirmed by
other experiments with comparable or larger
statistics~\cite{Abbasi:2004vu,Revenu:2005}, and probably a final
answer will come when the Pierre Auger Observatory~\cite{Auger}
will have collected enough data. Independently on the observation
of small-scale clustering, one could still look for large scale
anisotropies in the data, eventually correlating with some known
configuration of astrophysical source candidates. In this context,
the most natural scenario to be tested is that UHECRs correlate
with the luminous matter in the ``local'' universe. This is
particularly expected for candidates like gamma ray bursts (hosted
more likely in star formation regions) or colliding galaxies, but
it's also a sufficiently generic hypothesis to deserve an interest
of its own.

To this aim we use the  IRAS PSCz~\cite{saunders00a} astronomical
catalogue as tracer of Large Scale Structures from which, once
that the related selection effects have been taken into account,
we derive the related pattern of anisotropies; we then assess the
minimum statistics needed to detect the model against the
isotropic null hypothesis, in particular providing a forecast for
the Auger experiment. Previous attempts to address a similar issue
can be found
in~\cite{Waxman:1996hp,Smialkowski:2002by,Singh:2003xr}. Further
details of the analysis summarized here can be found
in~\cite{Cuoco:2005yd}.

%%%%%%%%%%%%%%%%%%%%%%%%%%%%%%%%%%%%%%%%%%%%%%%%%%%%%%%%%%%%%%%%%%%%%%%
\section{Data and Formalism}\label{astrodata}
%%%%%%%%%%%%%%%%%%%%%%%%%%%%%%%%%%%%%%%%%%%%%%%%%%%%%%%%%%%%%%%%%%%%%%%
%%%%%%%%%%%%%%%%%%%%%%%%%%%%
\subsection{The Catalogue}\label{cat}
%%%%%%%%%%%%%%%%%%%%%%%%%%%%

The IRAS PSCz catalogue~\cite{saunders00a} contains about 15.000
galaxies and related redshifts with a well understood completeness
function down to $z \sim 0.1$ (i.e. down to a redshift which is
comparable to the attenuation length introduced by the GZK effect)
and a sky coverage of about 84\%. The incomplete sky coverage is
mainly due to the so called zone of avoidance centered on the
galactic plane and caused by galactic extinction and to a few,
narrow stripes which were not observed with enough sensitivity by
the IRAS satellite (see~\cite{saunders00a}). These regions are
excluded from our analysis with the use of the binary mask
available with the PSCz catalogue itself. The mask $\mu$ is
defined so that $\mu(\hat{\Omega})=0,1$ if the direction
$\hat{\Omega}$ falls respectively inside or outside the blind
region.

To correctly employ the catalogue, selection and discreteness
systematic effects have to be considered. For the flux selection
correction the relevant quantity to be taken into account is the
fraction of galaxies actually observed at the various redshifts, a
quantity also known as the \emph{redshift selection function}
$\phi(z)$; given $\phi(z)$, the quantity $n(z)/\phi(z)$ represents
the experimental distribution corrected for the selection effects,
which must be used in the computations. The discreteness effects
(or "shot noise") are, instead, minimized using the catalogue only
until a maximum redshift of $z=0.06$ (corresponding to $180
\hMpc$), a fair compromise for which we have still good statistics
while keeping the intrinsic statistical fluctuations under
control. In any case, due to GZK effect in the energy range $E
\geq\,$5$\times 10^{19}\,$eV, the contribution from sources beyond
$z \simeq 0.06$ is sub-dominant, thus allowing to assume for the
objects beyond $z=0.06$ an effective isotropic source
contribution. A detailed discussion of the various catalogue
issues can be found in~\cite{Cuoco:2005yd} and references therein.

%%%%%%%%%%%%%%%%%%%%%%%%%%%%%%%%%%%%%%%%%%%%%%%%%%%%%%%%%%%%%%%%%%%%%%%
\subsection{Analysis}\label{method}
%%%%%%%%%%%%%%%%%%%%%%%%%%%%%%%%%%%%%%%%%%%%%%%%%%%%%%%%%%%%%%%%%%%%%%%
The goal of our analysis is to obtain the underlying probability
distribution $f_{\rm LSS}(\hat{\Omega},E)$ to have a UHECR with
energy higher than $E$ from the direction $\hat{\Omega}$.

For simplicity here we shall assume that each source of our
catalogue has the same probability to emit a UHECR, according to a
power law spectrum at the source $E^{-s}$; however, both
assumpitions can be easily generalized within the same
formalism~\cite{Singh:2003xr}.

A suitable expression for the distribution $f_{\rm
LSS}(\hat{\Omega},E)$ can be easily obtained as
\begin{eqnarray}
    f_{\rm LSS}(\hat{\Omega},E_{\rm cut}) \! \! \! \!   &  \propto  & \! \! \! \!  \sum_k \frac{1}{\phi(z_k)}
    \frac{\delta (\hat{\Omega}-\hat{\Omega}_k)}{4\pi d^2_L(z_k)}
\int_{E_i(E_{\rm cut},z_k)}^{\infty}  \! \! \! \! \! \! \! \! \!
\! \! \!
   E^{-s} {\rm d}E\nonumber  \\
     \! \! \! \!    & = & \! \! \! \!   \sum_k f_{\rm LSS}(k) \ \delta (\hat{\Omega}-\hat{\Omega}_k),
\label{spikemap}
\end{eqnarray}
that can be effectively seen as if at every source $k$ of the
catalogue it is assigned a weight $f_{\rm LSS}(k)$ that takes into
account geometrical effects through the luminosity distance
($d_L^{-2}$), selection effects ($\phi^{-1}$), and physics of
energy losses through the integral in d$E$; in this ``GZK
integral" the upper limit of integration is taken to be infinite,
though the result is practically independent on the upper cut used
provided it is much larger than 10$^{20}$ eV; the energy losses
are taken into account through the code described
in~\cite{Cuoco:2005yd} and parameterized in the function
$E_i(E_f,z)$, the ``initial energy function" giving the energy
$E_i$ of particle that should be injected at a redshift $z$ to
reach the Earth with an energy $E_f$.

We choose  $E_{\rm cut}=5 \times 10^{19}\,$eV that results in a
fair compromise between the intensity of the anisotropies (that
increases with energy) and the achievable statistics; for this
$E_{\rm cut}$ the isotropic contribution to the flux is
sub-dominant; however we can take it exactly into account and the
weight of the isotropic part is given by $ w_{\rm iso} \propto
\int_{z_{\rm GZK}}^{\infty} {\rm d}z\, p(z,E_{\rm cut})$
\footnote{the normalization factor is fixed consistently with
equation (\ref{spikemap})}.

Finally, to graphically represent the result, the spike-like map
of Eq. (\ref{spikemap}) is effectively smoothed through a gaussian
filter with beamwidth $\sigma=3^{\circ}$.

%%%%%%%%%%%%%%%%%%%%%%%%%%%%%%%%%%%%%%%%%%%%%%%%%%%%%%%%%%%%%%%%%%

Given the extremely poor statistics available in UHECR
astrophysics, we limit ourselves to address the basic issue of
determining the minimum number of events needed to significatively
reject ``the null hypothesis''. To this purpose, it is well known
that a $\chi^2$-test is an extremely good estimator and has no
ambiguity due to the 2-dimensional nature of the problem respect
to the K-S test or the Smirnov-Cramer-von Mises test. A criterion
guiding in the choice of the bin size is the following: with $N$
UHECRs events available and $M$ bins, one would expect ${\cal
O}(N/M)$ events per bin; to allow a reliable application of the
$\chi^2$-test, one has to impose $N/M\geq 10$. Each cell should
then cover at least a solid angle of $\Delta_M\sim
10\times\Delta_{\rm tot}/N$, $\Delta_{\rm tot}$ being the solid
angle accessible to the experiment. For $\Delta_{\rm tot}\sim
2\pi$ (50\% of full sky coverage), one estimates a square window
of side $454^{\circ}/\sqrt{N}$, i.e. $45^\circ$ for 100 events,
$14^\circ$ for 1000 events. Since the former number is of the
order of present world statistics, and the latter is the
achievement expected by Auger in a few years of operations, a
binning in windows of size $15^\circ$ represents quite a
reasonable choice for our forecast. This choice is also suggested
by the typical size of the observable structures, as can be seen
in the next section. Notice that the GMF, that induces at these
energies typical deflections of about $4^\circ$ \cite{KST05}, can
be safely neglected for this kind of analysis. The same remark
holds for the angular resolution of the experiment.

\begin{figure}[t]
\vspace{-4pc}
\includegraphics[angle=0,width=0.45\textwidth]{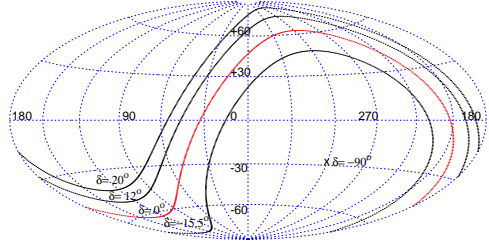}
%\centering \psfig{figure=fig4.eps,width=0.80\textwidth,angle=0}
\vspace{-2pc} \caption{\label{fig:RefFrame} Galactic coordinate
reference frame and contours enclosing 68\%, 95\% and 99\% of the
Auger exposure function, with the corresponding declinations. The
celestial equator ($\delta=0^\circ$) and south pole
($\delta=0,-90^{\circ}$) are also shown.}
\end{figure}

Moreover, for a specific experimental set-up one must include the
proper exposure $\omega_{\rm exp}$, that has to be convolved with
the previously found $f_{\rm LSS}$. We properly parameterize and
take into account the Auger exposure $\omega_{\rm exp}(\delta)$ as
function of declination $\delta$~\cite{Cuoco:2005yd,sommers2001}:
contour plots in galactic coordinates are shown in
Fig.~\ref{fig:RefFrame}.

For a given experiment and catalogue, the null hypothesis we want
to test is that the events observed are sampled---apart from a
trivial geometrical factor---according to the distribution $f_{\rm
LSS}\,\omega_{\rm exp}\,\mu$. Since we are performing a forecast
analysis, we will consider test realizations of $N$ events sampled
according to a random distribution on the (accessible) sphere,
i.e. according to $\omega_{\rm exp}\,\mu$, and determine the
confidence level (C.L.) with which the hypothesis is rejected as a
function of $N$. For each realization of $N$ events we calculate
the two functions
\begin{eqnarray}
{\cal X}_{\rm
iso}^2(N)=\frac{1}{M-1}\sum_{i=1}^{M}\frac{(o_i-\epsilon_i[f_{\rm
iso}])^2}{\epsilon_i[f_{\rm iso}]},\\
{\cal X}_{\rm
LSS}^2(N)=\frac{1}{M-1}\sum_{i=1}^{M}\frac{(o_i-\epsilon_i[f_{\rm
LSS}])^2}{\epsilon_i[f_{\rm LSS}]},
\end{eqnarray}
where $o_i$ is the number of ``random" counts in the $i$-th bin
$\Omega_i$, and $\epsilon_i[f_{\rm LSS}]$ and $\epsilon_i[f_{\rm
iso}]$ are the theoretically expected number of events in
$\Omega_i$ respectively for the LSS and isotropic distribution.
The mock data set is then sampled ${\cal N}$ times in order to
establish empirically the distributions of ${\cal X}_{\rm LSS}^2$
and ${\cal X}_{\rm iso}^2$,  and the resulting distribution is
studied as function of $N$ (plus eventually $s,E_{\rm cut}$,
etc.).

%%%%%%%%%%%%%%%%%%%%%%%%%%%%%%%%%%%%%%%%%%%%%%%%%%%%%%%%%%%%%%%%%%%%%%%
\section{Results}\label{results}
\begin{figure}[p]
\begin{center}
\begin{tabular}{c}
\includegraphics[angle=0,width=0.45\textwidth]{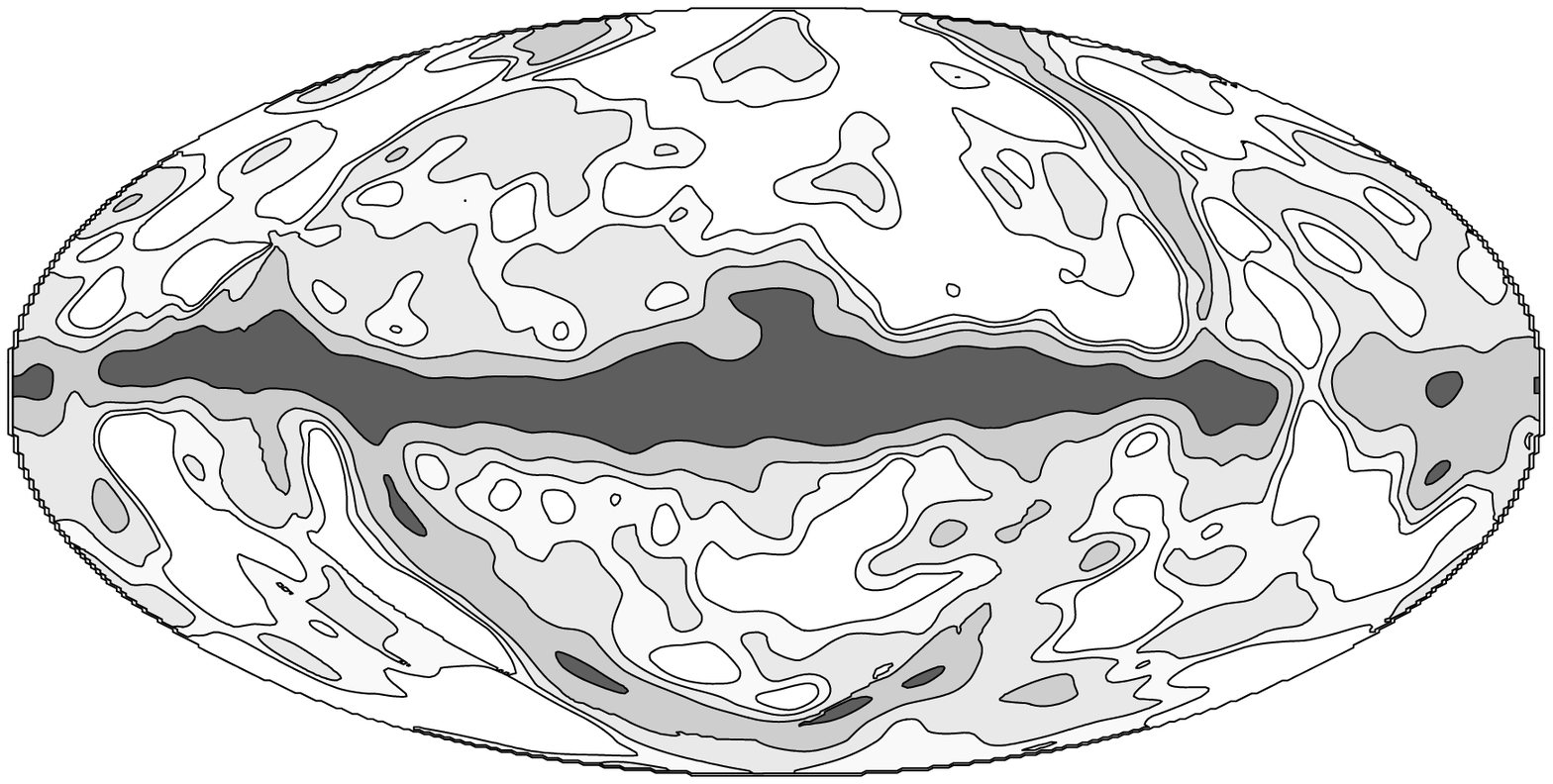}  \\
\includegraphics[angle=0,width=0.45\textwidth]{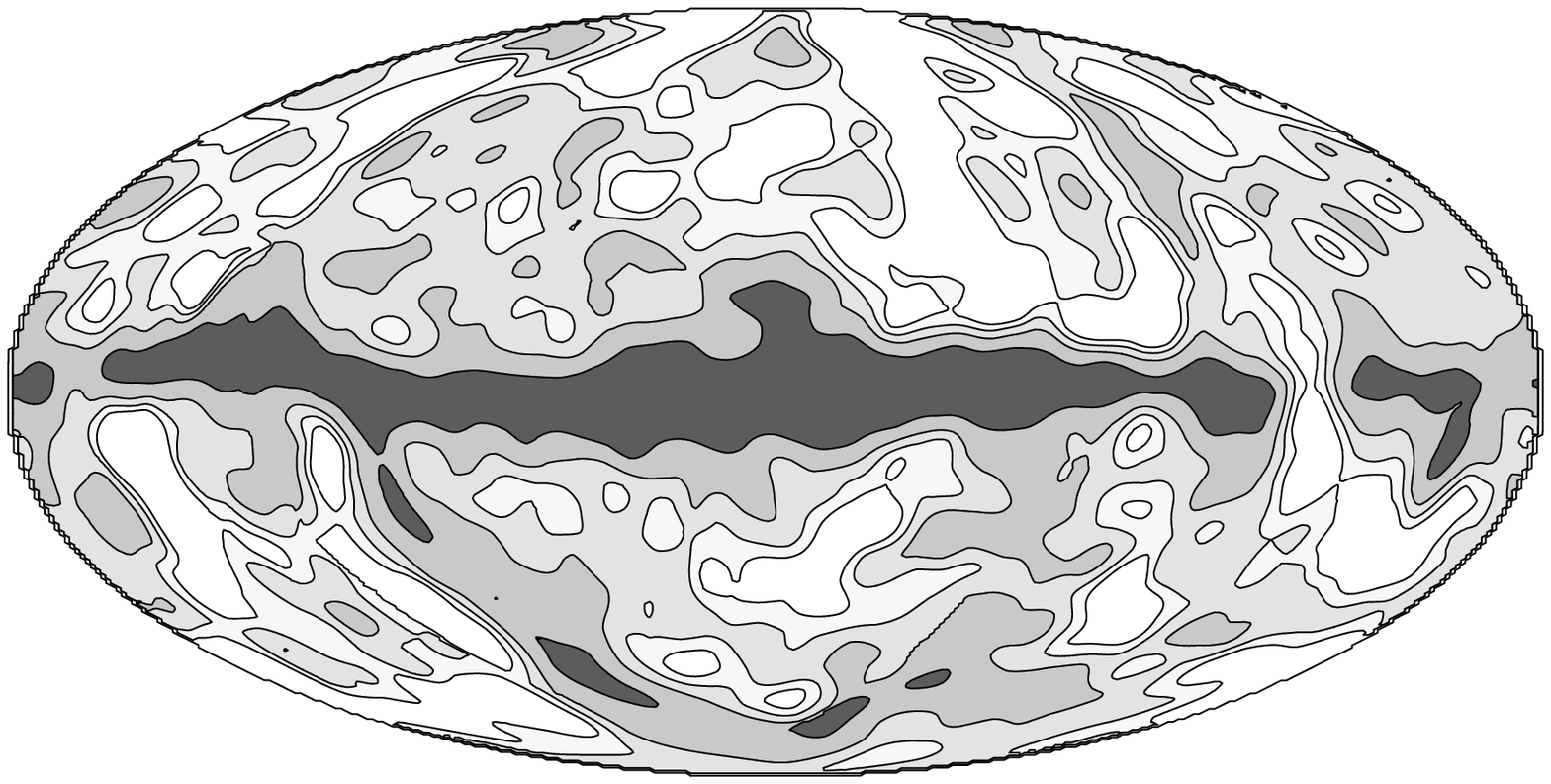}  \\
\includegraphics[angle=0,width=0.45\textwidth]{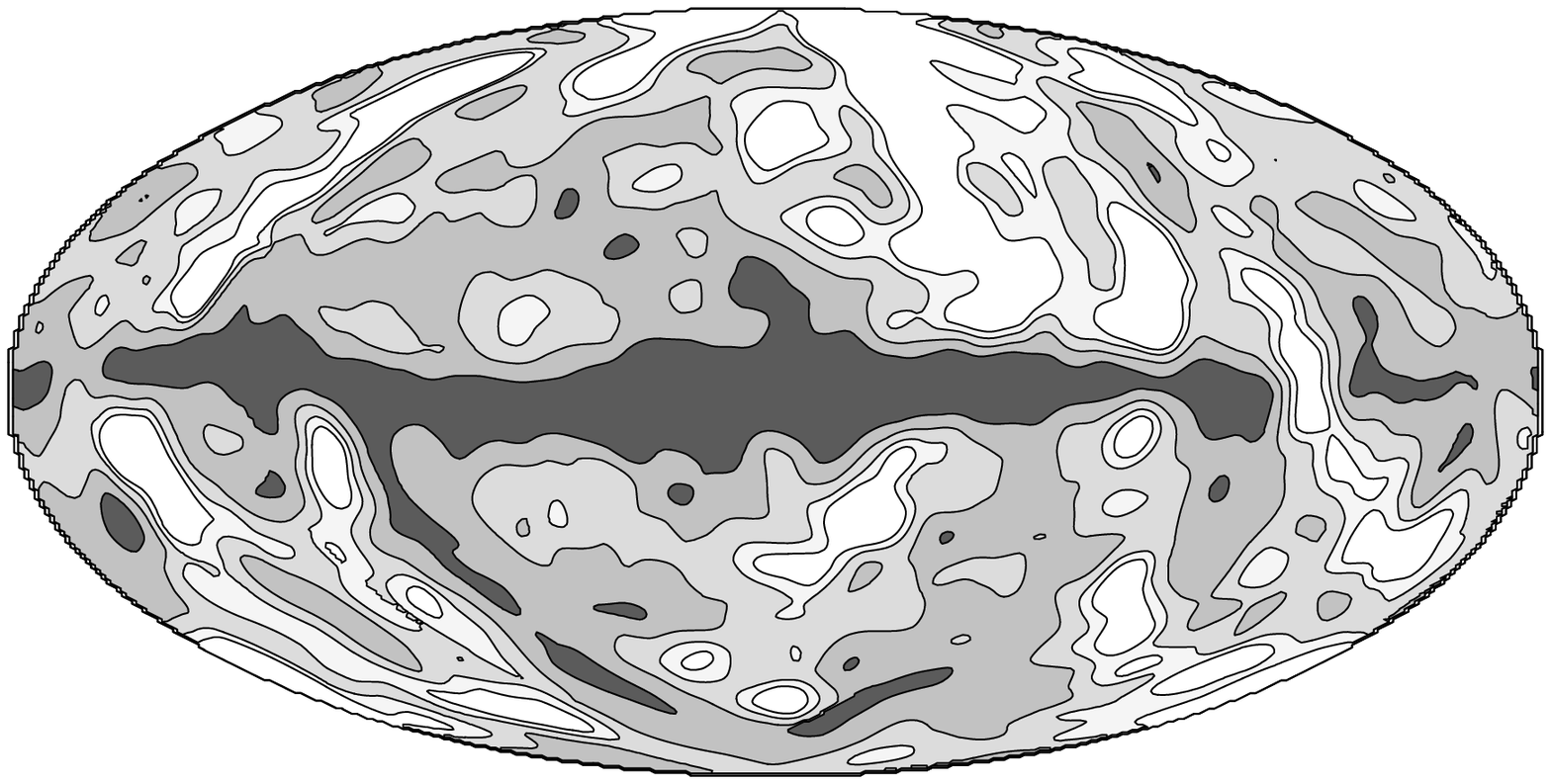}  \\
\includegraphics[angle=0,width=0.45\textwidth]{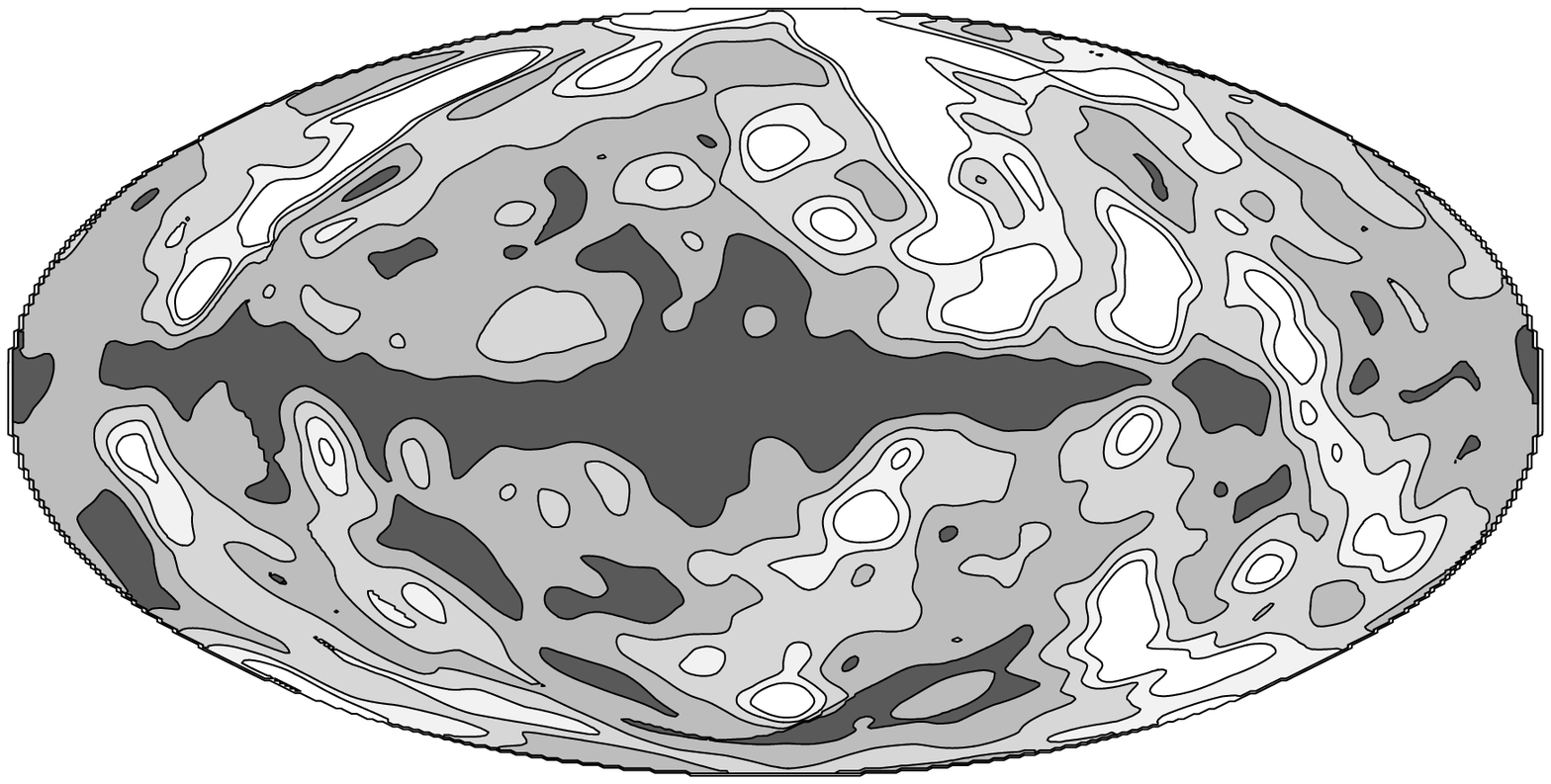}  \\
\end{tabular}
\end{center}
\caption{\label{fig:ArrayMapsEcut} Equal area Hammer-Aitoff
projections of the smoothed UHECRs arrival directions distribution
(Eq.~(\ref{spikemap})) in galactic coordinates obtained for fixed
$s=2.0$ and, from the upper to the lower panel, for $E_{\rm
cut}=3,5,7,9 \times 10^{19}$ eV. The smoothing angle is
$\sigma=3^{\circ}$. The contours enclose 95\%, 68\%, 38\%, 20\% of
the corresponding distribution. The dark central band corresponds
to the avoidance zone of the galactic plane.}
\end{figure}

%%%%%%%%%%%%%%%%%%%%%%%%%%%%%%%%%%%%%%%%%%%%%%%%%%%%%%%%%%%%%%%%%%%%%%%
In Fig.~\ref{fig:ArrayMapsEcut} we plot the smoothed maps in
galactic coordinates of the expected integrated flux of UHECRs
above the energy threshold $E_{\rm cut}=3,5,7,9\times 10^{19}\,$eV
and for slope parameter $s=2.0$; the isotropic part has been taken
into account and the ratio of the isotropic to anisotropic part
$w_{\rm iso}/\sum_k f_{\rm LSS}(k)$ is respectively
$83\%,3.6\%,\ll 1\%,\ll 1\%$.

Only for $E_{\rm cut}=3\times 10^{19}\,$eV the isotropic
background constitutes then a relevant fraction, since the GZK
suppression of far sources is not yet present. For the case of
interest $E_{\rm cut}=5\times 10^{19}\,$eV the contribution of
$w_{\rm iso}$ is almost negligible, while it practically
disappears for $E_{\rm cut}\agt 7\times 10^{19}\,$eV. Varying the
slope for $s=1.5,2.0,2.5,3.0$ while keeping $E_{\rm cut}= 5\times
10^{19}\,$eV fixed produces respectively the relative weights
$8.0\%,3.6\%,1.8\%,0.9\%$, so that only for very hard spectra
$w_{\rm iso}$ would play a non-negligible role.

\begin{figure}[t]
\includegraphics[angle=0,width=0.45\textwidth]{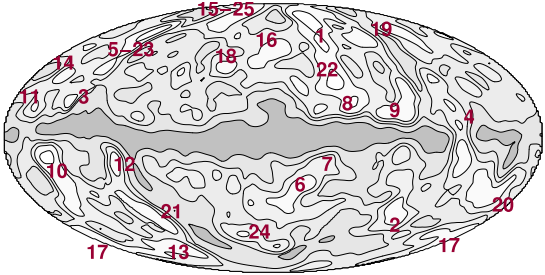}
\vspace{-1pc} \caption{\label{fig:legenda} Detailed key of the
structures visible in the UHECR maps; arbitrary contour levels.
Labels correspond to: (1) Southern extension of Virgo and Local
Supercluster; (2)Fornax-Eridani Cluster; (3) Cassiopea Cluster;
(4) Puppis Cluster; (5) Ursa Major Cloud; (6-7) Pavo-Indus and
"Great Attractor" region; (8) Centaurus Super-Cluster; (9) Hydra
Super-Cluster; (10) Perseus Super-Cluster; (11) Abell 569; (12)
Pegasus Cluster; (13-17) Pisces Cluster; (14) Abell 634; (15) Coma
Cluster;  (16-18) Hercules Supercluster; (19) Leo Supercluster;
(20) Columba Cluster; (21) Cetus Cluster; (22) Shapley
Concentration; (23) Ursa Major Supercluster; (24) Sculptor
Supercluster; (25) Bootes Supercluster.}
\end{figure}

Due to the GZK-effect, as it was expected, the nearest structures
are also the most prominent features in the maps. The most
relevant structure present in every slide is the Local
Supercluster. It extends along $l \simeq 140^{\circ}$ and $l
\simeq 300^{\circ}$ and includes the Virgo cluster at
$l=284^{\circ},b=+75^{\circ}$ and the Ursa Major cloud at
$l=145^{\circ},b=+65^{\circ}$, both located at $z \simeq 0.01$.
The lack of structures at latitudes from $l \simeq 0^{\circ}$ to
$l \simeq 120^{\circ}$ corresponds to the Local Void. At higher
redshifts the main contributions come from the Perseus-Pisces
supercluster ($l=160^{\circ},b=-20^{\circ}$) and the Pavo-Indus
supercluster ($l=340^{\circ},b=-40^{\circ}$), both at $z \sim
0.02$, and the very massive Shapley Concentration
($l=250^{\circ},b=+20^{\circ}$) at $z\sim 0.05$. For a more
detailed list of features in the map, see the key in
Fig.~\ref{fig:legenda}.

%%%%%%%%%%%%%
%
%
%%%%%%%%%%%%%

The $E_{\rm cut}$-dependence is clearly evident in the maps: as
expected, increasing $E_{\rm cut}$ results in a map that closely
reflects the very local universe (up to $z\sim0.03-0.04$) and its
large anisotropy; conversely, for $E_{\rm cut}\simeq3,4\times
10^{19}\,$eV, the resulting flux is quite isotropic and the
structures emerge as fluctuations from a background, since the GZK
suppression is not yet effective. This can be seen also comparing
the near structures with the most distant ones in the catalogue:
while the Local Supercluster is well visible in all slides, the
signal from the Perseus-Pisces super-cluster and the Shapley
concentration is of comparable intensity only in the two top
panels, while becoming highly attenuated for $E_{\rm cut}=7\times
10^{19}\,$eV, and almost vanishing for $E_{\rm cut}=9\times
10^{19}\,$eV. A similar trend is observed for increasing $s$ at
fixed $E_{\rm cut}$, though the dependence is almost one order of
magnitude weaker.

Looking at the contour levels in the maps we can have a precise
idea of the absolute intensity of the ``fluctuations'' induced by
the LSS; in particular, for the case of interest of $E_{\rm
cut}=5\times 10^{19}\,$eV the structures emerge only at the level
of 20\%-30\% of the total flux, the 68\% of the flux actually
enclosing almost all the sky. For $E_{\rm cut}=7,9\times
10^{19}\,$eV, on the contrary, the local structures are
significantly more pronounced, but in this case we have to face
with the low statistics available at this energies. Then in a
low-statistics regime it's not an easy task to disentangle the LSS
and the isotropic distributions.

\begin{figure}[t]
\begin{center}
\begin{tabular}{c}
\includegraphics[angle=0,width=0.45\textwidth]{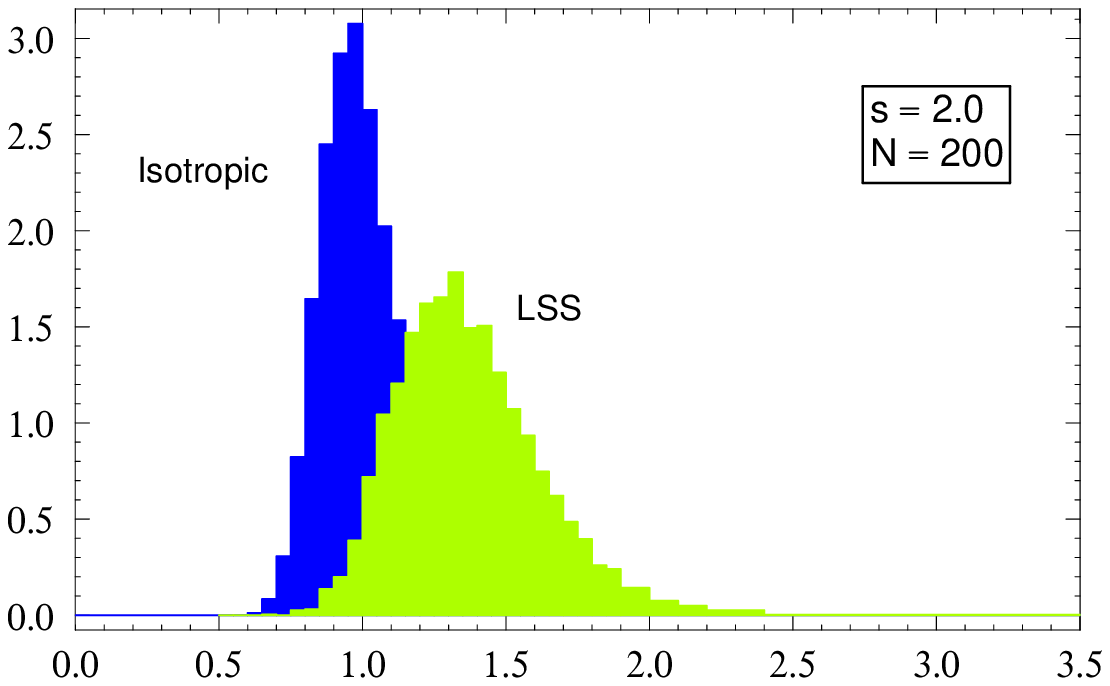} \\
\includegraphics[angle=0,width=0.45\textwidth]{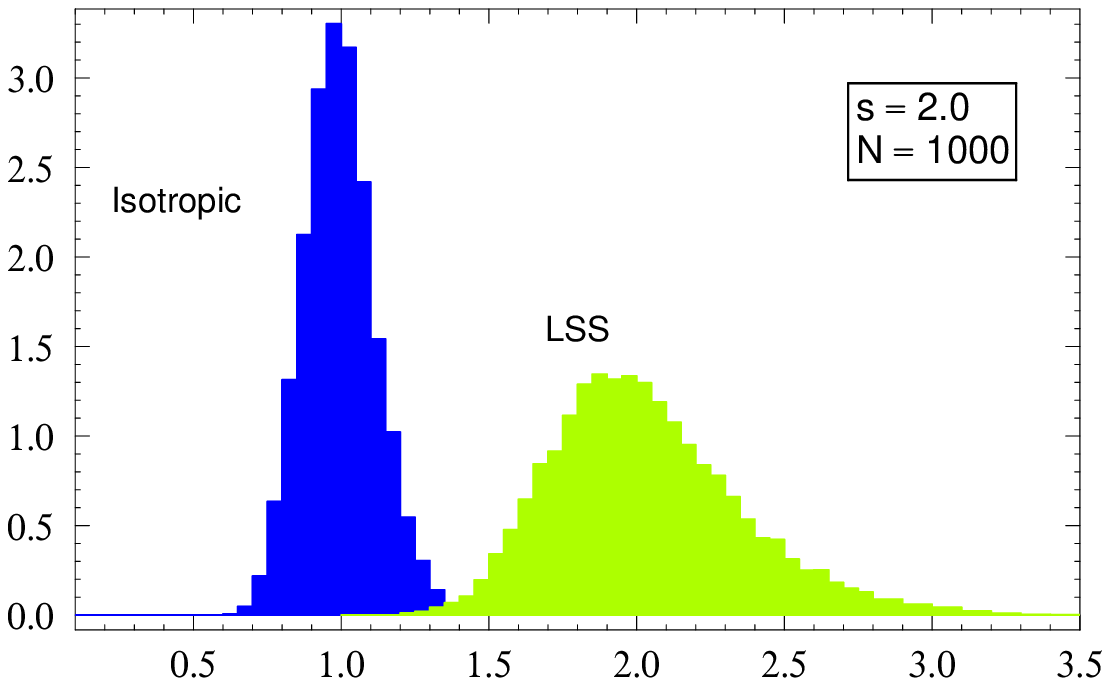}
\end{tabular}
\end{center}
 \vspace{-2pc}
\caption{\label{fig:chi2_nu} The distributions of the estimators
${\cal X}_{\rm iso}^2$ and ${\cal X}_{\rm LSS}^2$ for the cases
$s=2.0$ and for $N=200,1000$ events. The distribution are the
results of 10.000 monte-carlo simulation like described in the
text.} \vspace{-1pc}
\end{figure}

In Figure~\ref{fig:chi2_nu} we show the distributions of the
functions ${\cal X}_{\rm iso}^2$ and ${\cal X}_{\rm LSS}^2$
introduced in the previous section for $s=2.0$ and $N=200,1000$.
It is clear that a few hundreds events are hardly enough to
reliably distinguish the two models, while $N=\,$800--1000 should
be more than enough to reject the hypothesis at 2-3$\,\sigma$,
independently of the injection spectrum. Steeper spectra however
slightly reduce the number of events needed for a given C.L.
discrimination. Respect to the available data our results clearly
show that AGASA statistics (only 32 data at $E\geq 5\times
10^{19}\,$eV in the published data set~\cite{AGASA}, some of which
falling inside the mask) is too limited to draw any firm
conclusion on the hypothesis considered.
%%%%%%%%%%%%%%%%%%%%%%%%%%%%%%%%%%%%%%%%%%%%%%%%%%%%%%%%%%%%%%%%%%%%%%%
\section{Angular Power Spectrum}\label{power}
%%%%%%%%%%%%%%%%%%%%%%%%%%%%%%%%%%%%%%%%%%%%%%%%%%%%%%%%%%%%%%%%%%%%%%%
\begin{figure}[t]
\vspace{-1pc}
\begin{center}
\begin{tabular}{c}
\hspace{-3pc}
\includegraphics[angle=0,width=0.55\textwidth]{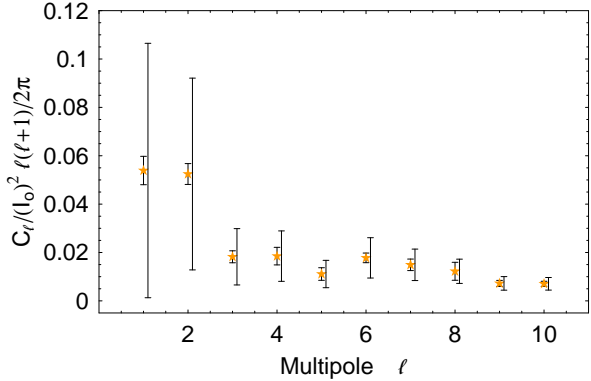}
\end{tabular}
\end{center}
\vspace{-3pc} \caption{\label{fig:power} The first multipoles of
the dimensionless angular power spectrum of the $5 \times 10^{19}$
eV map from the fitting procedure; also showed, by comparison, the
cosmic variance errors slightly shifted for clarity.}
\vspace{-1pc}
\end{figure}

The previous forecast analysis is completely independent from an
harmonic analysis of the anisotropies; however, a multipole
decomposition and the related power spectrum calculation greatly
helps in the understanding of the UHECRs anisotropies and, indeed,
a dipole anisotropy study has been the aim of various works
\cite{Mollerach:2005dv,Deligny:2004dj,Kachelriess:2006aq}; we
cover the gap in this section.

In the following we will concentrate on the $E_{\rm cut}=5 \times
10^{19}$ eV map; moreover, Auger is not expected to be very
sensitive to multipoles higher than~10, even after ten years of
operations, so we limit the analysis to the first few $l$.

\begin{figure}[t]
\begin{center}
\begin{tabular}{c}
\includegraphics[angle=0,width=0.42\textwidth]{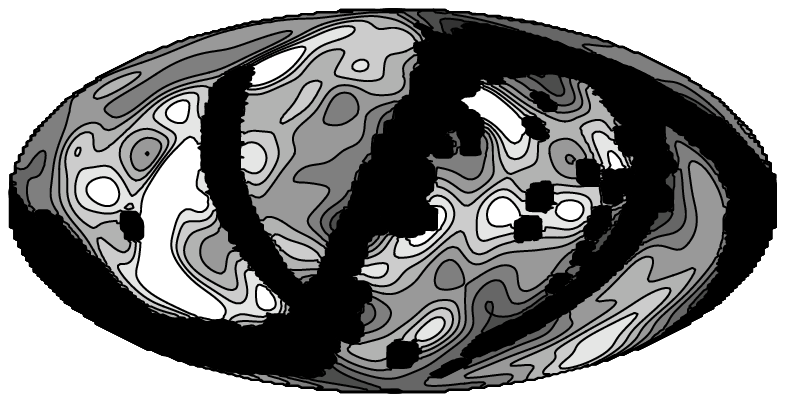} \\
\includegraphics[angle=0,width=0.42\textwidth]{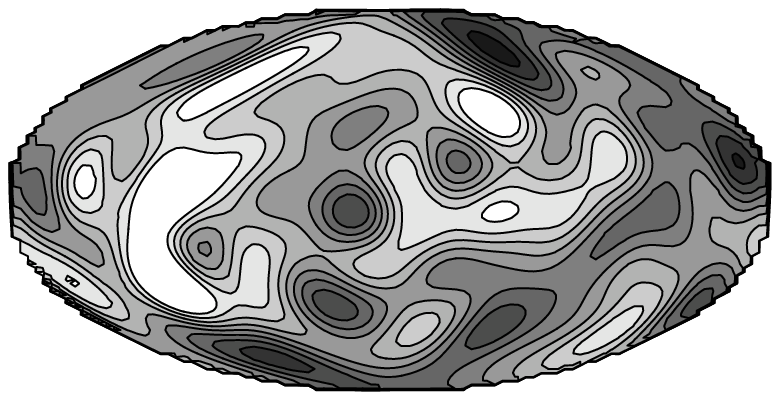} \\
\includegraphics[angle=0,width=0.42\textwidth]{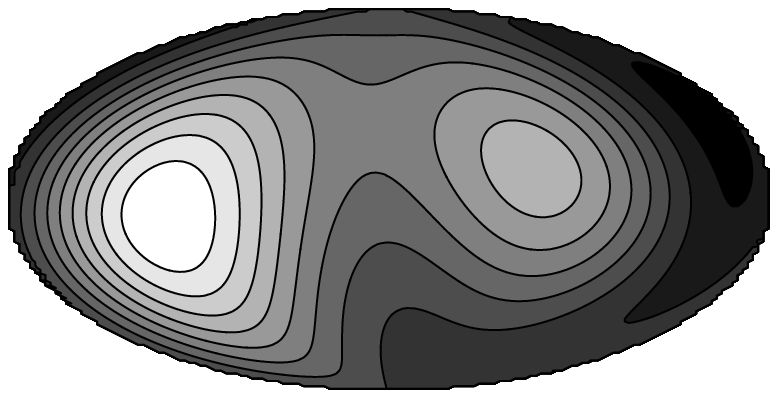}
\end{tabular}
\end{center}
\vspace{-2pc} \caption{\label{fig:multipoles} (a) The starting
UHECRs map and mask in equatorial coordinates. (b) The
reconstructed map from the first 10 multipoles. (c) As above for
dipole+quadrupole. } \vspace{-1pc}
\end{figure}

The main problem in calculating the Spherical Harmonics expansion
of the UHECRs map is the presence of the galactic avoidance zone;
however, in the first few $l$'s the typical angular scale
associated to the given multipole (of the order of $\sim \pi/l$)
is larger of the average angular extension of the galactic cut so
that an extrapolation in this zone is justified. We implement this
method fitting to the UHECRs map the sum of the first harmonics
till $l_{\rm max}$. The corresponding reconstructed map adding up
to the first $l$=10 and $l$=2 harmonics is showed in
Fig.~\ref{fig:multipoles} together with the original masked map.

Of course, filling the mask requires an extrapolation from the
near regions and this is somewhat questionable in relation to the
fact that some unknown cluster can be hidden behind our galaxy
changing in principle all the multipoles pattern; a not well
quantifiable error is indeed related to this ignorance though some
more sophisticated astronomical analysis indicate that it is
unlikely that relevant structures hide behind the galactic plane
\cite{Rowan-Robinson:1999mx,Lahav:1993ig,Lahav:1993igbis} so that
at this angular resolution ($l=10$ corresponds to $180/10\approx
20^{\circ}$ resolution) an extrapolation is quite justified.

In this respect the situation is quite different from the CMB case
where the anisotropies are isotropic and gaussian so that even a
limited region of sky represents a fair statistical sample of the
whole CMB sky, while the various multipoles are independent
realization of the underlying random field. The UHECRs field is
instead related to local and non-linear structures and it is then
far from being gaussian; moreover the presence of possible unknown
clusters in the avoidance zone introduces a correlation among the
various mutipoles so that the naive error bars derived from the
fit are not really trustable. We instead quantify the error
looking at the spread of the best fit values obtained varying the
degrees of freedom involved into the fit, related in its turn to
the chosen $l_{\rm max}$. We varied $l_{\rm max}$ in the range
1-12 that we found to be optimal.

The resulting angular power spectrum for the first 10 multipoles
of the adimensional map $f_{\rm LSS}(\hat{\Omega},E_{\rm cut}) / <
\! \! {f}_{\rm LSS} \! \! >$ is shown in Fig.~\ref{fig:power};
note that the true errors are not affected by the cosmic variance
given that we know the complete pattern of the angular
anisotropies and not only their statistical properties. The
hypothetical errors from the cosmic variance are, however, showed
by comparison: as expected the true errors on the very wide
structures of the dipole and quadrupole are quite small compared
to cosmic variance; the errors, then, rapidly degrade when the
scale related to the given $l$ approaches the size of the galactic
cut so that already for $l= 8-10$ the errors are almost comparable
to the cosmic variance. As expected the dipole gives a prominent
contribution though a similar power is predicted for the
quadrupole; this is easily explained in terms of the original map
where besides the Super-Galactic plane there is an important
contribution from the P-P cluster, which induces a mixed
dipolar-quadrupolar pattern. This contribution is almost an order
of magnitude greater respect to the other multipoles so that,
likely, it is also the dominant contribution in assessing the
minimum statistics needed to detect the anisotropies like
determined in the previous paragraphs. Since the dipole and
quadrupole are the easiest anisotropies to look for in UHECR
analyses, this result suggests that the $l$=1+$l$=2 pattern should
slowly emerge from the cosmic rays data as the first detectable
large scale anisotropy. Given the relevant role represented by the
quadrupole and dipole we give in table~\ref{table2} the expression
of the $a_{lm}$ and the related errors, in equatorial coordinates.
Finally, it is worth to notice the scale of the multipole
anisotropies of the order $\sim 0.06$ that translates into an
overall anisotropy of the order $\sim \sqrt{0.06}\approx 25\%$ in
good agreement with what found with analysis of the full map.
\begin{table}
\begin{center}
\begin{tabular}{|c|cc|}
\hline
$a_{lm}$ & Re & Im\\
\hline
$a_{10}$   & 0.008 $\pm$ 0.027 &  ...\\
$a_{11}$  & 0.230 $\pm$ 0.014 & 0.148 $\pm$ 0.039 \\
$a_{20}$  &  -0.226 $\pm$ 0.049 & ...\\
$a_{21}$  & -0.072 $\pm$ 0.035 &  0.093 $\pm$ 0.044 \\
$a_{22}$  &  0.251 $\pm$ 0.042  & -0.136 $\pm$ 0.021 \\
\hline
\end{tabular}
\caption{Low $l$ multipole moments in equatorial coordinates.
$a_{l-m}$ moments are obtained through the reality condition
$a_{l-m}=a_{lm}^*$.
  \label{table2}}
\end{center}
\vspace{-2pc}
\end{table}

As check of the method we perform an independent estimate of the
power spectrum using the Master algorithm \cite{Hivon:2001jp}, a
method typically employed in similar CMB analysis. This method is
well suited for small scale anisotropies and is quite
complementary to the fitting procedure; we verified that the two
spectra superimpose and join smoothly making us confident on the
reliability of the result.

%%%%%%%%%%%%%%%%%%%%%%%%%%%%%%%%%%%%%%%%%%%%%%%%%%%%%%%%%%%%%%%%%%%%%%%
\section{Summary and conclusion}\label{concl}
%%%%%%%%%%%%%%%%%%%%%%%%%%%%%%%%%%%%%%%%%%%%%%%%%%%%%%%%%%%%%%%%%%%%%%%
UHECRs offer the fascinating possibility to make astronomy with
high energy charged particles; in particular, besides small scale
clustering, a large scale anisotropy is expected correlating with
the local cosmological LSS. However, we showed with a $\chi^2$
approach that a huge statistics of several hundreds data is
required to start testing the model at Auger South.

Untill now, the lack of UHECR statistics has seriously limited the
usefulness of such a kind of analysis. However, progresses are
expected in forthcoming years. The Southern site of Auger is
already taking data and, once completed, the total area covered
will be of 3000 km$^2$, thus improving by one order of magnitude
present statistics in a couple of years~\cite{Bertou:2005bx}. The
idea to build a Northern Auger site strongly depends on the
possibility to perform UHECR astronomy, for which full sky
coverage is of primary importance. To this aim, the
Japanese-American Telescope Array in the desert of Utah is
expected to become operational by 2007~\cite{Arai:2003dr} offering
almost an order of magnitude larger aperture per year than AGASA
in the Northern sky.

%%%%%%%%%%%%%%%%%%%%%%%%%%%%%%%%%%%%%%%%%%%%%%%%%%%%%%%%%%%%%%%%%%%%%%%
%\section*{References}
%%%%%%%%%%%%%%%%%%%%%%%%%%%%%%%%%%%%%%%%%%%%%%%%%%%%%%%%%%%%%%%%%%%%%%%

\end{document}